\documentclass{ws-ijmpcs}
\usepackage{slashed}
\def\trimmarks{}

\allowdisplaybreaks[4]
\begin{document}

\markboth{S. Y. Wei}
{Higher twist effects in $e^+e^-$ annihilation at high energies}

%
\catchline{}{}{}{}{}
%

\title{Higher twist effects in $e^+e^-$ annihilation at high energies}

\author{Shu-yi Wei}

\address{School of Physics, Shandong University,\\
Jinan, Shandong, 250100,
China\\
shuyi@mail.sdu.edu.cn}



\maketitle

\begin{history}
\received{Day Month Year}
\revised{Day Month Year}
\published{Day Month Year}
\end{history}

\begin{abstract}
In the two papers published recently\cite{Wei:2013csa,Wei:2014pma}, 
we apply collinear expansion to both inclusive ($e^+ e^- \to h+X$) 
and semi-inclusive ($e^+ e^- \to h + \bar q + X$) hadron production 
in $e^+ e^-$ annihilation to derive a formalism suitable for 
a systematic study of leading as well as higher twist contributions to 
fragmentation functions at the tree level.
We carry out the calculations for hadrons with spin-0, spin-1/2 as well as spin-1. 
This proceeding is mainly a summary of these two papers.
\end{abstract}

\ccode{PACS numbers: 13.66.Bc, 13.87.Fh, 13.88.+e, 12.15.Ji, 12.38.-t, 12.39.St, 13.40.-f, 13.85.Ni}

\section{Introduction}


The $e^+ e^-$ annihilation process is most suitable to study fragmentation functions among all different high energy reactions, as there is no hadron involved in the initial state.
The one dimensional fragmentation functions can be studied via inclusive process, while, the 3D information can only be extracted from semi-inclusive process.

Higher twist terms may have very important contributions to azimuthal asymmetries and spin asymmetries\cite{Qiu:1991pp},
which are usually measured in experiments to study the properties of fragmentation functions and parton distribution functions.
Collinear expansion, first developed in 1980s, seems to be the unique method to calculate leading twist and higher twist contributions in a systemic way.

This method was applied to inclusive DIS process\cite{Ellis:1982wd} to get the cross section up to twist-4 level at first.
It was summarized as four steps\cite{Liang:2006wp}, and was applied to SIDIS to get the form of azimuthal asymmetries up to twist-3\cite{Liang:2006wp} and twist-4\cite{Song:2010pf}. 
Recently, we applied collinear expansion to inclusive $e^+ e^-$ annihilation process\cite{Wei:2013csa} and semi-inclusive process\cite{Wei:2014pma}.

\section{Collinear Expansion}

Collinear expansion was inspired by the collinear approximation,
which is stated as follows.
\begin{itemize}
\item We only keep the collinear component of the quark momentum, $k_i\approx p/z_i$.
\item We only keep the plus component of the gluon field, $A^{\mu} \approx A^+ \bar n^\mu$.
\end{itemize}
This approximation is reasonable, if we are only care about the leading twist contributions, since other components are power suppressed compared to the plus component. 
We apply this approximation to the following diagrams, and we have,
\begin{figure}[h!]\centering
\includegraphics[width=0.9\textwidth]{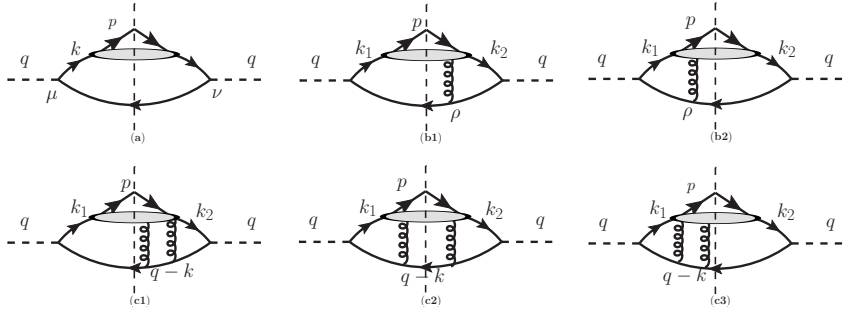}
\caption{The first few Feynman diagrams as examples of the diagram series with exchange of $j$ gluon(s).}
\end{figure}
\begin{align}
\tilde{W}_{\mu\nu}^{\rm approx} = \frac{1}{2p^+} \mathrm{Tr}\left[ \Gamma_\mu \slashed n \Gamma_\nu  \hat \Xi^{\rm approx} (z_B,p,S,k'_\perp) \right],
\end{align}
where,  $\Gamma_\mu=\gamma_\mu (c_V^q  - c_A^q \gamma_5)$ is the vertex for weak interaction.
And the soft matrix $\Xi^{\rm approx}$ is a little different for inclusive process and semi-inclusive process. For the inclusive process, this soft matrix does not dependent on $k'_\perp$,
\begin{align}
\hat \Xi^{\rm approx}_{\rm in} (z,p,S)= \hat{\mathcal{F}} (\xi^-) \sum_X \langle 0 | \mathcal{L}^\dagger(0^-,\infty)\psi(0) |{ hX}\rangle \langle { hX}| \bar\psi(\xi^-) \mathcal{L}(\xi^-,\infty) |0\rangle.
\end{align}
While, for the semi-inclusive process, 
\begin{align}
\hat \Xi^{\rm approx}_{\rm si} (z, k'_{\perp},p,S) = \hat{\mathcal{F}} (\xi^-, \vec{\xi}_\perp)  \sum_X  \langle 0| \mathcal{L}^\dagger (0,\infty) \psi(0)|{ hX\rangle \langle hX}| \bar\psi(\xi) \mathcal{L} (\xi,\infty) |0 \rangle .
\end{align}
$\hat{\mathcal{F}}$ is just the Fourier Transformation operator.
We see, these soft matrices are automatically gauge invariant.

If we are going to the twist-3 level, these transverse components can not be simply neglected any more.
So, we need the collinear expansion to take them back into account by the following steps\cite{Wei:2013csa,Wei:2014pma}.
\begin{itemize}
\item Make a Taylor expansion to the hard part, $H^{i}(k) = H^{i} (k=p/z) + \left. \frac{\partial H}{\partial k}\right|_{k=p/z} \times (k-p/z) +\cdots$.
\item Decompose the gluon fields in this way, $A^\rho (y) = A^+(y) \bar n^\rho + (A^\rho (y) - A^+(y) \bar n^\rho)$. 
\end{itemize}
Reducing them with Ward identities, we get the hadronic tensor twist by twist,
\begin{align}
W_{\mu\nu} = \tilde W_{\mu\nu}^{(0)} +  \tilde W_{\mu\nu}^{(1L)} + ( \tilde W_{\mu\nu}^{(1L)} )^* + \cdots,
\end{align}
where,
\begin{align}
& \tilde W_{\mu\nu}^{(0)} = \frac{1}{2} {\rm Tr} \left[ \hat {h}^{(0)}_{\mu\nu} ~  \hat \Xi^{(0)}  \right], 
& & \tilde W_{\mu\nu}^{(1L)} = - \frac{1}{4p\cdot q} {\rm Tr} \left[ \hat {h}^{(1)\rho}_{\mu\nu}~ \omega_\rho^{\ \rho'} \hat \Xi_{\rho'}^{(1)} \right].
\end{align}
$\tilde W^{(0)}$ is, actually, $\tilde{W}_{\mu\nu}^{\rm approx}$, and $\hat \Xi^{(0)} = \hat \Xi^{\rm approx} $.
$\tilde W_{\mu\nu}^{(1L)}$ is the new hadronic tensor, whose leading contribution is at twist-3.
The new quark gluon correlators are defined as,
\begin{align}
&\hat \Xi^{(1)~\rm in}_{\rho} = \hat{\mathcal{F}} (\xi^-) \sum_X \langle 0| \mathcal{L}^\dagger (0, \infty) D_\rho (0) \psi (0) |hX \rangle \langle  { hX}| \bar\psi(\xi^-) \mathcal{L}(\xi^-,\infty) |0\rangle,\\
&\hat \Xi^{(1)~\rm si}_{\rho} = \hat{\mathcal{F}} (\xi^-,\vec{\xi}_\perp) \sum_X \langle 0| \mathcal{L}^\dagger (0, \infty) D_\rho (0) \psi (0) |hX \rangle \langle  { hX}| \bar\psi(\xi) \mathcal{L}(\xi,\infty) |0\rangle.
\end{align} 
These $\Xi$'s defined above are all  $4$ by $4$ matrices which can alway be decomposed in terms of Gamma matrices.
In this case, only $\gamma^\alpha$ and $\gamma_5 \gamma^\alpha$ that will contribute.
\begin{align}
\hat \Xi^{(i)} = \Xi^{(i)}_{\alpha} \gamma^\alpha + \tilde \Xi^{(i)}_\alpha \gamma_5 \gamma^\alpha,
\end{align}
where, parity invariant constrains the possible structures of them,
since $\Xi^{(i)}_\alpha$  is a vector and $\tilde \Xi^{(i)}_\alpha$ is an axis vector.
We will discuss the details of this decomposition in the following two sections for inclusive process and semi-inclusive process separately.

\section{Inclusive Process $e^+ e^- \to h + X$}

These $\Xi$'s are made up by $p$, $n$ and $S$ for the inclusive process.
$S$ here refers to the spin parameters of produced hadrons.
For spin-$1/2$ hadrons, we only need the spin vector $S^\mu$, 
while for vector mesons, we need also a spin tensor $T^{\mu\nu}$, which can be decomposed in terms of $S_{LL}$, $S_{LT}^\mu$ and $S_{TT}^{\mu\nu}$.


\subsection{Spin Independent Part and Spin Vector Dependent Part}
\begin{align}
& z \Xi^{(0)\alpha}_{\rm in} = p^\alpha \left[ D_1(z) + S_{LL} D_{1LL}(z) \right] +  
  M \left[\epsilon_\perp^{\alpha\gamma} S_{\perp\gamma} D_T(z) + S_{LT}^\alpha D_{LT}(z)\right]  ,\\
& z \tilde \Xi^{(0)\alpha}_{\rm in} = \lambda_h p^\alpha \Delta D_{1L}(z) +  M S_\perp^\alpha \Delta D_T(z)  + 
  M \epsilon_\perp^{\alpha\gamma} S_{LT,\gamma} \Delta D_{LT}(z),\\
& z \Xi^{(1)\rho\alpha}_{\rm in} =  p^\alpha\left[M \epsilon_\perp^{\rho\gamma} S_{\perp\gamma}  \xi_{\perp S}^{(1)}(z)+
  M S_{LT}^\rho  \xi_{LTS}^{(1)}(z)\right]  ,\\
& z \tilde \Xi^{(1)\rho\alpha}_{\rm in} = i p^\alpha \left[M S_\perp^\rho  \tilde \xi_{\perp S}^{(1)} (z) + 
 i  M \epsilon_\perp^{\rho\gamma} S_{LT,\gamma}   \tilde\xi_{LTS}^{(1)}(z)\right].
\end{align}

For spin-$0$ hadrons, only the spin independent terms that will contribute,
\begin{equation}
\frac{d^2\sigma^{unp}}{dzdy}= \sum_{q,c} \frac{2\pi\alpha^2}{Q^2} \chi  T_0^q (y) D_1^q (z) .
\end{equation}

For spin-$1/2$ particles, both the spin independent part and spin vector dependent part contribute. The spin vector dependent part is given by,
\begin{align}
\frac{d \sigma^{Vpol}}{dzdy} = \frac{2\pi\alpha^2}{Q^2} \chi \Bigl\{ 
\lambda_{ h} T_1 (y) \Delta D_{1L} (z)&  +\frac{4M}{z Q^2} 
\bigl[ T_2(y) D_T(z) \epsilon_\perp^{l_\perp S_\perp} \nonumber\\
&+ T_3(y)\Delta D_T(z) l_\perp \cdot S_\perp \bigr]\Bigr\}.
\end{align}
Here, the summation over quark flavor and color is not written out explicitly.
We see, the first term is at leading twist and  the last two terms are at twist-3.
So, it is obvious that at leading twist, produced hadrons will be longitudinally polarized.
\begin{equation}
P_{Lh}=\frac{\sum_{q} T_{1}^{q} (y)\Delta D_{1L}^{q\to h}(z)}{\sum_{q} T_{0}^{q}(y)D_1^{q\to h}(z)}=\frac{\sum_{q} P_{q}(y) T_{0}^{q}(y)\Delta D_{1L}^{q\to h}(z)}{\sum_{q} T_{0}^{q}(y)D_1^{q\to h}(z)}.
\end{equation}
It is just proportional to the polarization of quark, 
$P_{q}(y)=T_{1}^{q}(y)/T_{0}^{q}(y)$, 
times a spin transfer, $\Delta D_{1L}^{q\to h}(z)$.

There are also two transverse polarizations at twist-3.
One is perpendicular to the leptonic plane, $\epsilon_\perp^{l_\perp S_\perp}$, while the other one lies in the leptonic plane, $l_\perp \cdot S_\perp$.
\begin{align}
&P_{hy}= \frac{4M}{zQ} \frac{\sum_{q} \tilde T_{2}^{q}(y) D_T^{q\to h}(z)}{\sum_{q} T_{0}^{q} (y)D_1^{q\to h}(z)},
&&P_{hx}=-\frac{4M}{zQ} \frac{\sum_{q} \tilde T_{3}^{q}(y) \Delta D_T^{q\to h}(z)}{ \sum_{q} T_{0}^{q}(y)D_1^{q\to h}(z)}.
\end{align}
The first one is T-odd. Hence this is NO corresponding term in inclusive DIS process.
The last one is P-odd, which will disappear in the electromagnetic process.

\subsection{Spin Tensor Dependent Part}

\begin{align}
& z \Xi^{(0)\alpha}_{\rm in}  = p^\alpha S_{LL} D_{1LL}(z) + M S_{LT}^\alpha D_{LT}(z) , 
~~~ z \tilde \Xi^{(0)\alpha}_{\rm in} =  
  M \varepsilon_\perp^{\alpha\gamma} S_{LT,\gamma} \Delta D_{LT}(z),\\
& z \Xi^{(1)\rho\alpha}_{\rm in}  =  p^\alpha  M S_{LT}^\rho  \xi_{LTS}^{(1)}(z), 
~~~~~~~~~~~~~~~~~~ z \tilde \Xi^{(1)\rho\alpha}_{\rm in} = i  M p^\alpha \varepsilon_\perp^{\rho\gamma} S_{LT,\gamma}   \tilde\xi_{LTS}^{(1)}(z).
\end{align}

For vector mesons, the cross section contains contributions from spin independent, vector polarization dependent and tensor polarization dependent parts.
\begin{align}
\frac{d\sigma^{Tpol}}{dzdy} = \frac{2\pi\alpha^2}{Q^2} \chi
\Bigl\{  T_0(y) S_{LL}  D_{1LL}(z) & + \frac{4M}{zQ^2} \bigl[  T_2 (y)  l_\perp \cdot S_{LT} D_{LT}(z)
 \nonumber\\
&+ T_3 (y) \epsilon_\perp^{l_\perp S_{LT}} \Delta D_{LT}(z) \bigr] \Bigr\}.
\end{align}
The most interesting spin parameter of vector mesons is spin alignment ($\rho_{00}$).
This leading twist effect is verified by LEP experiment.
We predict that this tensor polarization can also be easily measured by BES experiment since it is P-even.
\begin{align}
& \rho_{00}^{\rm weak}= \frac{1}{3} - \frac{1}{3} \frac{\sum_{q} T_{0}^{q}(y) D_{1LL}^{q\to h}(z) }{\sum_{q} T_{0}^{q}(y) D_1^{q\to h}(z) }, &&
\rho_{00}^{\rm em} = \frac{1}{3} - \frac{1}{3} \frac{\sum_{q} e_q^2D_{1LL}^{q\to h}(z)}{\sum_{q} e_q^2D_1^{q\to h}(z)}.
\end{align}

\section{Semi-inclusive Process $e^+ e^- \to h + \bar q + X$ }

For the semi-inclusive process, these $\Xi$'s are also 
functions of the transverse momentum $k'_\perp$.
So, we have much more new structures.

\subsection{Spin Independent Part and Azimuthal Asymmetries}
\begin{align}
&z\Xi^{(0)}_\alpha =p_\alpha \hat D_1 +k_{\perp \alpha} \hat D^\perp,
&&z\tilde \Xi^{(0)}_\alpha  =\epsilon_{\perp \alpha k_\perp} \Delta \hat D^\perp ,\\
& z \Xi^{(1)}_{\rho \alpha} = p_\alpha k_{\perp \rho} \xi^{(1)}_\perp, 
&& z \tilde \Xi^{(1)}_{\rho\alpha} = i p_\alpha\epsilon_{\perp \rho k_\perp} \tilde \xi^{(1)}_\perp.
\end{align}

And the corresponding cross section is,
\begin{align}
&\frac{d\sigma^{(si,unp)}}{dydzd^2k'_\perp}=\frac{\alpha^2\chi}{2\pi Q^2}\Big\{T_0^q (y) \hat D_1 
+ \frac{4}{zQ^2} \big[T_2^q (y) l_\perp\cdot k'_\perp \hat D^\perp + 
T_3^q (y) \epsilon_\perp^{l_\perp k'_\perp} \Delta \hat D^\perp \big] \Big\}. 
\end{align}
We see immediately that there are two azimuthal asymmetries at twist-3,
\begin{align}
&A^{\cos\varphi}_{unp}=
 -\frac{2 |\vec k'_\perp|}{z Q} \frac{\sum_q \tilde T_2^q (y) \hat D^{\perp q\to h}}{\sum_q T_0^q (y) \hat D_1^{q\to h} }, &&
A^{\sin\varphi}_{unp}=
 \frac{2 |\vec k'_\perp|}{z Q} \frac{\sum_q \tilde T_3^q (y) \Delta \hat D^{\perp q\to h}}{\sum_q T_0^q (y) \hat D_1^{q\to h}}.
\end{align}
The second one is P-odd, so it will disappear in electromagnetic process.

\subsection{Spin Vector Dependent Part}
\begin{align}
z\Xi^{(0)}_\alpha  &  =
 p_\alpha \frac{\epsilon_{\perp}^{k_\perp S_\perp}}{M}  \hat D_{1T}^\perp 
+ k_{\perp\alpha} \frac{\epsilon_{\perp}^{k_\perp S_\perp} }{M}  \hat D_{T}^\perp  + \lambda_h \epsilon_{\perp \alpha k_\perp} \hat D_L^\perp 
+ M \epsilon_{\perp \alpha S_\perp} \hat D_T ,  \\
z\tilde \Xi^{(0)}_\alpha  &  =
p_\alpha \Big[ \lambda_h \Delta \hat D_{1L} 
 + \frac{k_\perp \cdot S_\perp}{M} \Delta \hat D_{1T}^\perp \Big]  + \frac{\epsilon_{\perp}^{k_\perp S_\perp}}{M}  \epsilon_{\perp \alpha k_\perp} \Delta \hat D_T^\perp \nonumber\\
& +  \lambda_h k_{\perp\alpha}\Delta \hat D_L^\perp + M S_{\perp \alpha} \Delta \hat D_T, \\
z\Xi^{(1)}_{\rho\alpha}  & =  p_\alpha \Bigl[
 M \epsilon_{\perp \rho S_\perp} \xi^{(1)}_{T} + k_{\perp\rho}\frac{\epsilon_{\perp}^{k_\perp S_\perp}}{M}   \xi^{(1)\perp }_{T}
+  \lambda_h \epsilon_{\perp \rho k_\perp} \xi^{(1)\perp}_{L} \Bigr], \\
z\tilde \Xi^{(1)}_{\rho \alpha} &  =
i p_\alpha \Bigl[M S_{\perp \rho} \tilde \xi^{(1)}_{T} +  \frac{\epsilon_{\perp}^{k_\perp S_\perp} }{M} \epsilon_{\perp \rho k_\perp}\tilde \xi^{(1)\perp }_{T} 
+  \lambda_h k_{\perp\rho} \tilde \xi^{(1)\perp}_{ L}\Bigr].
\end{align}
The cross section is then given by,
\begin{align}
 &\frac{d\sigma^{(si,Vpol)}}{dydz d^2k'_\perp}=\frac{\alpha^2\chi }{2\pi Q^2} 
 \Bigl\{  T_0^q (y) \frac{\epsilon_\perp^{k'_\perp S_\perp}}{M}  \hat D_{1T}^\perp +T_1^q (y) \big[ \lambda_h \Delta \hat D_{1L} + \frac{k'_\perp\cdot S_\perp}{M} \Delta \hat D_{1T}^\perp \big]\nonumber\\
&+\frac{4\lambda_h}{zQ^2} \big[ T_2^q (y)  \epsilon_\perp^{l_\perp k'_\perp}  \hat D_L^\perp + 
T_3^q (y) l_\perp \cdot k'_\perp  \Delta \hat D_L^\perp \big] +\frac{4\epsilon_\perp^{k'_\perp S_\perp} }{zMQ^2} \big[ T_2^q (y) l_\perp \cdot k'_\perp \hat D_T^\perp \nonumber \\
& + T_3^q (y) \epsilon_\perp^{l_\perp k'_\perp} \Delta \hat D_T^\perp \big] +\frac{4M}{zQ^2} \big[ T_2^q (y)  \epsilon_\perp^{l_\perp S_\perp}  \hat D_T + 
T_3^q (y) l_\perp \cdot S_\perp  \Delta \hat D_T \big] \Bigr\}.
\end{align}
We see, that there is a leading twist polarization in the longitudinal direction, 
\begin{align}
 P_{Lh}^{(0)} = 
\frac{\sum_q T_0^q (y) P_q(y)\Delta \hat D_{1L} }{\sum_q T_0^q(y) \hat D_1 }.
\end{align}
The most suitable directions to study transverse polarization and corresponding fragmentation functions are those in and transverse to the production plane. At leading twist,
\begin{align}
& P_{hn}^{(0)} = - \frac{|\vec k'_\perp|}{M}
\frac{\sum_q T_0^q (y)  \hat D_{1T}^\perp }{\sum_q T_0^q(y) \hat D_1 }, 
&& P_{ht}^{(0)} = - \frac{|\vec k'_\perp|}{M}
\frac{\sum_q  P_q (y)T_0^q(y) \Delta \hat D_{1T}^\perp }{\sum_q T_0^q(y) \hat D_1 }.
\end{align}
Those twist-3 terms gives us corrections\cite{Wei:2014pma} which are suppressed by $M/Q$.

\subsection{Spin Alignment}

The Lorentz structures for vector mesons are much more complicated, 
since there are much more spin parameters.
We would not show you all those structures in this proceeding, for the limitation of length.
Please find the details in [2] if you are interested in this topic.
Here we only show you results concerning spin alignment.
\begin{align}
 &z\Xi^{(0)}_{\alpha LL} = p_\alpha S_{LL}  \hat D_{1LL}+ k_{\perp\alpha}  S_{LL} \hat D_{LL}^\perp, 
& &z\tilde \Xi^{(0)}_{\alpha LL} = \epsilon_{\perp\alpha k_\perp}  S_{LL} \Delta \hat D_{LL}^\perp, \\
 &z\Xi^{(1)}_{\rho\alpha LL} =p_\alpha k_{\perp\rho} S_{LL} \xi_{LL}^\perp ,
& &z\tilde \Xi^{(1)}_{\rho\alpha LL} = i p^\alpha\epsilon_{\perp\rho k_\perp}  S_{LL}  \tilde \xi_{LL}^\perp.
\end{align}
The corresponding cross section is given by,
\begin{align}
\frac{d\sigma^{(si,LL)}}{dydz d^2k'_\perp}=\frac{\alpha^2\chi}{2\pi Q^2} S_{LL} &
\Big\{ T_0^q (y)\hat D_{1LL} + \frac{4}{zQ^2} \big[ T_2^q (y) l_\perp\cdot k'_\perp \hat D_{LL}^\perp + T_3^q (y) \epsilon_{\perp }^{l_\perp {k'_\perp}} \Delta \hat D_{LL}^\perp\big]\Big\}.
\end{align}
And then, we get the spin alignment up to twist-3,
\begin{align}
S_{LL} &=\frac{\sum_q T_0^q(y)\hat D_{1LL}}{2\sum_q T_0^q(y) \hat D_1}\Bigl[1+\frac{M}{Q}\hat\Delta\Bigr]
- \frac{\sum_q 2 \Bigl[\tilde T_2^q (y) k'_x \hat D_{LL}^\perp  - \tilde T_3^q (y) k'_y \Delta \hat D_{LL}^\perp \Bigr] }{zQ \sum_q T_0^q (y) \hat D_1}.
\end{align}

\section{Summary}

In inclusive process,
there is a leading twist longitudinal polarization for spin-$\frac{1}{2}$ hadrons and also spin alignment ($\rho_{00}\neq \frac{1}{3}$) for vector mesons.
At twist-3, there are transverse polarizations for spin-$\frac{1}{2}$ hadrons in and transverse to the leptonic plane.

In semi-inclusive process,
for spin-0 hadrons, there are two azimuthal asymmetries at twist-3. 
For spin-$\frac{1}{2}$ hadrons, there is a longitudinal polarization and also transverse polarizations  in and transverse to the production plane at leading twist.


The author would like to thank all my collaborators, Zuo-tang Liang, Yu-kun Song and Kai-bao Chen. This work was supported in part by the National Natural Science Foundation of China (project No. 11035003 and 11375104), and the Major State Basic Research Development Program in China (No. 2014CB845406).


\begin{thebibliography}{0}    



\bibitem{Wei:2013csa} 
  S.~-y.~Wei, Y.~-k.~Song and Z.~-t.~Liang,
  Phys.\ Rev.\ D {\bf 89}, 014024 (2014).

\bibitem{Wei:2014pma} 
  S.~-y.~Wei, K.~-b.~Chen, Y.~-k.~Song and Z.~-t.~Liang,
  arXiv:1410.4314 [hep-ph].



\bibitem{Qiu:1991pp} 
  J.~-w.~Qiu and G.~F.~Sterman,
  Phys.\ Rev.\ Lett.\  {\bf 67}, 2264 (1991);
  Phys.\ Rev.\ D {\bf 59}, 014004 (1999); and 
  R.~L.~Jaffe and X.~-D.~Ji,
  Phys.\ Rev.\ D {\bf 43}, 724 (1991).


\bibitem{Ellis:1982wd} 
  R.~K.~Ellis, W.~Furmanski and R.~Petronzio,
  Nucl.\ Phys.\ B {\bf 207}, 1 (1982); 
%
%
  Nucl.\ Phys.\ B {\bf 212}, 29 (1983).
  

\bibitem{Liang:2006wp} 
  Z.~-t.~Liang and X.~-N.~Wang,
  Phys.\ Rev.\ D {\bf 75}, 094002 (2007)


\bibitem{Song:2010pf} 
  Y.~-k.~Song, J.~-h.~Gao, Z.~-t.~Liang and X.~-N.~Wang,
  Phys.\ Rev.\ D {\bf 83}, 054010 (2011).






\end{thebibliography}
\end{document}